\documentclass[aps,amsmath,amsfonts,10pt]{revtex4}
\usepackage{epsfig,graphicx}
\usepackage[english]{babel}
\usepackage{amsfonts}
\usepackage{amsmath}
\usepackage{latexsym}
\usepackage{graphics,bm}
\usepackage{subfigure}
\usepackage{dcolumn}
\usepackage{bm}
\usepackage{rotating}

\begin{document}

\title{Bogoliubov spectrum of a cigar shaped Fermi superfluid in an optical lattice at the BEC-BCS crossover}

\author{Aranya B Bhattacherjee}
\affiliation{Max Planck-Institute f\"ur Physik Komplexer Systeme, N\"othnitzer Str.38,01187 Dresden,Germany \\ and \\ Department of Physics, A.R.S.D College, University of Delhi (South Campus), New Delhi-110021, India. }

\begin{abstract}
We study the Bogoliubov spectrum of an elongated Fermi superfluid confined in a one-dimensional superfluid along the Bose-Einstein-condensate (BEC)-Bardeen-Cooper-Schrieffer (BCS) crossover. We derive analytic expressions for the effective mass and the Bogoliubov excitation spectrum of the axial quasiparticles along the crossover based on the hydrodynamic theory. Our investigation reveal interesting signatures of BEC-BCS crossover in an optical lattice which deserve experimental investigation. 
\end{abstract}

\maketitle

\section{Introduction}

The experimental realization \cite{Modugno03,Jochim03} of optical lattices for fermionic isotopes such as $^{6}Li$ or $^{40}K$ is stimulating new perspectives in the study of superfluidity in these systems. An increase in the superfluid transition temperature when using potentials created by standing light waves has been predicted \cite{Hofstter02}. It has been shown that for values of the Fermi energy above the first Bloch band, the center of mass motion of a Fermi gas trapped in an one-dimensional periodic potential is overdamped in the collisional regime due to Umklapp process \cite{Orso04}. A hydrodynamic theory of Fermi superfluids in the presence of optical lattice has been developed in the weakly interacting BCS limit \cite{Pitae05}.

 In such a fermionic system, it should be possible to adjust the interaction strength and light intensity to tune the system continuously between two limits: a Bardeen-Cooper-Schrieffer (BCS) type superfluid (involving correlated atom pairs in momentum space) and a Bose-Einstein condensate (BEC) in which spatially local pairs of atoms are bound together. This crossover between BCS-type superfluidity and the BEC limit for a dilute gas of fermionic atoms has been of recent theoretical interest \cite{Parish05}. In particular lot of theoretical attention has focussed on the collective excitations at the BEC-BCS crossover \cite{Ghosh06}. The first experimental results on the collective frequencies of the lowest axial and radial breathing modes of ultracold gases of $^{6}Li$ across the Feshbach resonance have also become available \cite{Bartenstein04}. In atomic Fermi gases, tunable strong interactions are produced using the Feshbach resonance \cite{Houbiers98}. Feshbach resonance, occurs when the energy of a quasibound molecular state becomes equal to the energy of two free atoms. The magnetic field dependence of the resonance allows precise tuning of the atom-atom interaction strength. Across the resonance the s-wave scattering length goes from large positive to large negative values. The fermionic system becomes molecular BEC for strongly repulsive interaction and transforms into a BCS superfluid when the interaction is attractive. Recent experiments have entered the crossover regime and yielded results of the interaction strength by the cloud size and expansion \cite{Hara02}. 

In atomic Fermi gas experiments, BEC-BCS crossover regime in optical lattices has not yet been demonstrated. Considering the fact that experiments in this area are making rapid progress, we were motivated to study for the first time the low energy Bogoliubov spectrum of a cigar shaped superfluid Fermi gas confined in an one-dimensional optical lattice along the BEC-BCS crossover regime using the hydrodynamic approach. For fermions confined in a trapping potential, the density profile changes slowly in space if the particle number of the system is large enough. Under such conditions, a local density approximation can be applied to the state and a hydrodynamic approach can be adopted to investigate the low energy collective modes. Expressions for the effective mass and various collective modes are new results of this work.

\section{The effective action and the hydrodynamic equations in an optical lattice}
We consider a cigar shaped dilute ultracold fermionic gas trapped in an one-dimensional optical lattice. The optical lattice is formed by two counterpropagating laser beams, for example in the $z$ direction.

\begin{equation}
V_{op}(z)=sE_{R}\sin^{2}\left(\frac{\pi z}{d} \right). 
\end{equation}

Here, $d$ is the lattice period and $s$ is the dimensionless amplitude of the lattice potential. $E_{R}=\dfrac{\hbar^2\pi^2}{2md^2}$ is the recoil energy ($\omega_{R}=\frac{E_{R}}{\hbar}$ is the corresponding recoil frequency) of the lattice. In addition to the optical lattice, we also have a harmonic trap $V_{ho}(r,z)=\dfrac{m}{2}\left(\omega^{2}_{r} r^{2}+\omega^{2}_{z} {z}^{2} \right)$. Mass of the fermionic atom is $m$. $\omega_{r}$ and $\omega_{z}$ are the radial and the axial trap frequencies. We take $\omega_{r}>\omega_{z}$. The harmonic oscillator frequency corresponding to small motion about the minima of the optical lattice is $\omega_{s}\approx \dfrac{\sqrt{s_{1}}\hbar \pi^2}{md^2}$.  $\omega_{s}>>\omega_{z}$ so that the optical lattice dominates the harmonic potential along the $z$-direction and hence the harmonic potential is neglected. The strong laser intensity will give rise to an array of several quasi-two dimensional pancake shaped condensates. In writing down the effective action, we will follow Wouters et al. \cite{Wouters04}. When an optical lattice is present along the $z$-direction, we can decouple the free motion in the $x,y$-plane from the tunneling motion in the $z$-direction. The partition function for a system consisting of layers of 2D fermions is

\begin{equation}
Z=\int {\cal D} \psi^{\dagger}_{j,\sigma}(r){\cal D} \psi_{j,\sigma}(r) exp\left(-S\left[\frac{\psi^{\dagger}_{j,\sigma}(r),\psi_{j,\sigma}(r)}{\hbar} \right]  \right), 
\end{equation}

where the action is given by

\begin{eqnarray}
S\left[ \psi^{\dagger}_{j,\sigma}(r),\psi_{j,\sigma}(r)\right]&& =\sum_{j}\sum_{\sigma}\int_{0}^{\hbar \beta} d\tau \int  d \mathbf {x} ~ \psi^{\dagger}_{j,\sigma}(r)\left(\hbar \partial_{\tau}-\frac{\hbar^{2} \nabla^{2}}{2m}+V_{ext}(j)-\mu \right)\psi_{j,\sigma}(r)- \nonumber \\&& U\psi^{\dagger}_{j,\uparrow}(r) \psi^{\dagger}_{j,\downarrow}(r)  \psi_{j,\downarrow}(r)\psi_{j,\uparrow}(r)+J\left( \psi^{\dagger}_{j,\sigma}(r)\psi_{j+1,\sigma}(r)+\psi^{\dagger}_{j+1,\sigma}(r)\psi_{j,\sigma}(r)\right). 
\end{eqnarray}

Here the three vector notation is used, $r=(x,y,\tau)$. $\beta=1/k_{B}T$, $T$ is the temperature. Also $\mathbf{x}=x,y$.  The field $\psi_{j,\sigma}(r)$ belongs to a fermion of mass $m$ in layer $j$ and spin $\sigma$($\uparrow$ or $\downarrow$). $U$ is the attractive strength between the fermions. The interlayer Josephson tunneling energy is $J=sE_{R}\left( \dfrac{\pi^{2}}{4}-1\right)exp\left( -\sqrt{s}\left( \dfrac{\pi}{2}\right)^{2} \right)$\cite{Martikainen03}. $V_{ext}(j)$ is the external potential acting on each layer and can be parabolic in addition to the optical lattice. In order to get rid of the quartic term in equation (3) one needs to perform a Hubbard-Stratonovic (HS) transformation and introduce the HS fields $\Delta_{j}(r)\propto <\psi_{j} \psi_{j}>$, $\theta_{j}(r)$, after which the integration over fermionic variables is performed. Our goal is an investigation of the superfluid properties of the ultracold Fermi system using the hydrodynamic approach. The hydrodynamic interpretation of $|\Delta_{j}(r)|^{2}$ is that it represents the density of fermion pairs, whereas $\vec v(r)=\dfrac{\hbar \nabla_{r}\theta_{j}(r)}{m}$ is interpreted as the superfluid velocity field. The Hubbard-Stratonovich transformation transforms equation (2) into:

\begin{equation}
Z=\int {\cal D}\psi _{j,\sigma }^{\dag }(r){\cal D}\psi _{j,\sigma }(r){\cal %
D}\Delta _{j}(r){\cal D}\Delta _{j}^{\dagger }(r)\exp \left\{
-S^{\left( 1\right) }/ \hbar \right\} ,
\end{equation}%
with%
\begin{eqnarray}
S^{\left( 1\right) } &=&\sum_{j}\int_{0}^{ \hbar \beta }d\tau \int d {\bf x}%
\left[ \frac{\left\vert \Delta _{j}\left( r\right) \right\vert ^{2}}{U}%
\right.  \nonumber \\
&&+\sum_{\sigma =\pm 1}\psi _{j,\sigma }^{\dag }\left( r\right) \left(
\hbar \partial _{\tau }-\hbar^{2} \frac{\nabla ^{2}}{2m}+V_{ext}\left( j\right) -\mu \right)
\psi _{j,\sigma }\left( r\right)  \nonumber \\
&&- \Delta _{j}\left( r\right) \psi _{j,\uparrow }^{\dag }\left(
r\right) \psi _{j,\downarrow }^{\dag }\left( r\right) - \Delta _{j}^{\dag
}\left( r\right) \psi _{j,\downarrow }\left( r\right) \psi _{j,\uparrow
}\left( r\right)  \nonumber \\
&&+\left. J\sum_{\sigma }\left( \psi _{j,\sigma }^{\dag }\left( r\right)
\psi _{j+1,\sigma }\left( r\right) +\psi _{j+1,\sigma }^{\dag }\left(
r\right) \psi _{j,\sigma }\left( r\right) \right) \right] .
\end{eqnarray}%

In order to have information about the physical density, we multiply the partition function by constant $C$ : 

\begin{eqnarray}
C &=&\int {\cal D}\zeta _{j}(r){\cal D}\rho _{j}(r)\exp \left\{
-\frac{1}{\hbar}\sum_{j}
\int_{0}^{ \hbar \beta }d\tau \int d {\bf x}\;i\zeta _{j} \left( r\right)
\right.  \nonumber \\
&&\left. \times \left[ \rho _{j}\left( r\right) -\psi _{j,\uparrow }^{\dag
}\left( r\right) \psi _{j,\uparrow }\left( r\right) -\psi _{j,\downarrow
}^{\dag }\left( r\right) \psi _{j,\downarrow }\left( r\right) \right]
\right\} .  \label{eqC}
\end{eqnarray}%

Carrying out the functional integral over the new HS field $\zeta _{j}\left( r\right) $
gives $\delta \lbrack \rho _{j}\left( r\right) -\psi _{j,\uparrow
}^{\dag }\left( r\right) \psi _{j,\uparrow }\left( r\right) -\psi
_{j,\downarrow }^{\dag }\left( r\right) \psi _{j,\downarrow }\left( r\right)
]$ and thus $\rho _{j}\left( r\right) $ corresponds to the physical density
of the system along any path. In order to investigate the BCS gap and the
phase we separate the complex field $\Delta _{j}(r)$ in a modulus and a
phase  $ \Delta _{j}\left( r\right) =\left| \Delta _{j}\left( r\right)
\right| e^{i\theta _{j}\left( r\right) }$ and also transform the fermion
fields as $\psi _{j,\sigma }\left( r\right) \rightarrow \psi _{j,\sigma
}\left( r\right) e^{i\theta _{j}\left( r\right) /2}$. Additionally, we shift
the field $i\zeta _{j}\left( r\right) $ according to 
\begin{equation}
i\zeta _{j}\left( r\right) \rightarrow i\zeta _{j}\left( r\right)
+i \hbar \partial _{\tau }\frac{\theta _{j}\left( r\right) }{2}+ \hbar^{2} \frac{\left( \nabla
\theta _{j}\left( r\right) \right) ^{2}}{8m}+V_{ext}\left( j\right) -\mu,
\end{equation}%
and use the Nambu spinor notation $\eta _{j}\left( r\right) =\left( \psi
_{j,\uparrow }\left( r\right) ,\psi _{j,\downarrow }^{\dag }\left( r\right)
\right) ^{T}$. 

After this procedure and a further path integral over the fermion fields results in the partition function to be written as 

\begin{equation}
Z\propto \int {\cal D}\left\vert \Delta _{j}(r)\right\vert {\cal D}%
\theta _{j}(r){\cal D}\zeta _{j}(r){\cal D}\rho _{j}(r)\exp \left\{ -S_{%
\text{eff}}/ \hbar \right\}, 
\end{equation}%
where 
\begin{equation}
S_{\text{eff}}=S_{0}+\hbar \text{Tr}\left[ \ln (-\tilde{G}^{-1}/ \hbar )\right]  \label{Seff},
\end{equation}%

\begin{eqnarray}
S_{0} &=&\sum_{j}\int_{0}^{ \hbar \beta }d\tau \int d{\bf x}\left\{ \frac{%
\left\vert \Delta _{j}(r)\right\vert ^{2}}{U}\right.  \nonumber \\
&&\left. +\left[ i  \zeta _{j}(r)+i  \partial _{\tau }\frac{\theta
_{j}\left( r \right) }{2}+  \frac{\left( \nabla \theta _{j}\left( r\right)
\right) ^{2}}{8m}+V_{ext}\left( j\right) -\mu \right] \rho _{j}\left(
r\right) \right\}, 
\end{eqnarray}%
and 
\begin{eqnarray}
-\tilde{G}^{-1}\left( r,j;r^{\prime },j^{\prime }\right) &=&\delta \left(
r-r^{\prime }\right) \left\{ \delta _{jj^{\prime }}\left[ \left( \hbar \partial
_{\tau }-i\frac{\hbar^{2} \nabla \theta _{j}\left( r\right) }{2m}\nabla -i\frac{ \hbar^{2} \nabla
^{2}\theta _{j}\left( r\right) }{4m}\right) \sigma _{0}\right. \right. 
\nonumber \\
&&\left. +\left( -\frac{ \hbar^{2} \nabla ^{2}}{2m}-i  \zeta _{j}(r)\right) \sigma
_{3}-  \left\vert \Delta _{j}(r)\right\vert \sigma _{1}\right]  \nonumber
\\
&&\left. +\delta _{j+1,j^{\prime }}J e^{i\left( \theta _{j+1}-\theta
_{j}\right) \sigma _{3}/2}\sigma _{3}+\delta _{j-1,j^{\prime
}}J e^{-i\left( \theta _{j+1}-\theta _{j}\right) \sigma _{3}/2}\sigma
_{3}\right\} .
\end{eqnarray}%

Here $\sigma_{i}$ are the Pauli matrices. Now to determine the low energy dynamics of the density and the phase of the superfluid, the paths along which $\theta_{j}(r)$ and $\rho_{j}(r)$ vary slowly in comparison to the fermionic frequencies (Fermi energy and binding energy) are of importance. Along these paths we have to make saddle point approximations for the fields $\left| \Delta _{j}(r)\right| $
and $\zeta _{j}(r)$ as:

\begin{eqnarray}
\left| \Delta _{j}(r)\right| &=&\left| \Delta _{j}^{(0)}\left( r\right)
\right| +\delta \left| \Delta _{j}(r)\right| , \\
\zeta _{j}(r) &=&\zeta _{j}^{(0)}\left( r\right) +\delta \zeta
_{j}(r),
\end{eqnarray}%
where $\left| \Delta _{j}^{(0)}\left( r\right) \right| $ and $\zeta
_{j}^{(0)}\left( r\right) $ also vary slowly in comparison to the fermion
frequencies. Since we are interested in the low-energy regime of $S_{eff}$, corrections that go beyond this regime are neglected \cite{Palo98}. Since we are dealing with $T=0$ dynamics, the normal component is absent and the superfluid is stiff. This means that we neglect terms proportional to $\dfrac{\nabla^{2} \theta_{j}(r)}{m}$, which is equivalent to $\vec \nabla .\vec v=0$ (we consider steady flow).

Expanding the effective action (equation 9) and the Green's function (equation 11) around the saddle point values $\left| \Delta _{j}^{(0)}\right| $ and $\zeta _{j}^{(0)}=-iz_{j}$ leads to the saddle point effective action as a sum of contributions independent of $J$ and tunneling contributions($S_{eff,j\rightarrow j+1}^{tunnel}$):

\begin{eqnarray}
S_{\text{eff}}^{\text{0}} &=&%
\hbar \mathop{\rm Tr}%
\left[ \ln \left( -G_{0}^{-1}/\hbar \right) \right] +\sum_{j}\int_{0}^{\hbar \beta }d\tau
\int d{\bf x} \left \{ \frac{\left| \Delta _{j}^{(0)}(r)\right| ^{2}}{U}%
\right.  \nonumber \\
&&\left. +\left[ z_{j}(r)+i \hbar \partial _{\tau }\frac{\theta _{j}\left( r\right) 
}{2}+ \hbar^{2} \frac{\left( \nabla \theta _{j}\left( r\right) \right) ^{2}}{8m}%
+V_{ext}\left( j\right) -\mu \right] \rho _{j}\left( r\right) + S_{eff,j \rightarrow j+1}^{tunnel}\right\},
\end{eqnarray}%

where,

\begin{equation}
-G_{0}^{-1}=\sigma_{0}\left(\hbar \frac{\partial}{\partial \tau} \right)+\sigma_{3}\left(-\frac{\hbar^{2}}{2m}\nabla^{2}-z_{j} \right)-\sigma_{1}\left(|\Delta_{j}^{0}| \right).   
\end{equation}

The saddle point equations are:

\begin{eqnarray}
\frac{1}{U} &=&\int \frac{d^{2}{\bf k}}{\left( 2\pi \right) ^{2}}\frac{%
1-2n_{F}\left[ E_{j}(k)\right] }{2E_{j}\left( k\right) },  \label{egap0} \\
\rho _{j}(r) &=&\int \frac{d^{2}{\bf k}}{\left( 2\pi \right) ^{2}}\left( 
\frac{\hbar^{2} k^{2}}{2m}-i\zeta _{j}^{(0)}(r)\right) \left\{ \frac{2n_{F}\left[
E_{j}(k)\right] -1}{E_{j}\left( k\right) }+1\right\} ,  \label{edens}
\end{eqnarray}%
with $n_{F}(E)=1/(e^{\beta E}+1)$ the Fermi-Dirac distribution function and $%
E_{j}\left( k\right) $ the local BCS energy defined by%
\begin{equation}
E_{j}\left( k\right) =\sqrt{\left( \frac{\hbar^{2} k^{2}}{2m}-i\zeta
_{j}^{(0)}(r)\right) +\left| \Delta _{j}^{(0)}(r)\right| ^{2}}.
\label{BCSenergy}
\end{equation}%

In the limit of $T=0$ and Thomas Fermi (TF) approximation (neglecting the kinetic energy), the local BCS energy is rewritten as 

\begin{equation}
E_{j}=\sqrt{z_{j}^{2}+\Delta^{2}}.
\end{equation}

Here, we have used the saddle point value in the absence of vortex $|\Delta_{j}^{0}(r)|=\Delta$. The first saddle point equation (equation 16) corresponds to the BCS gap equation, whereas the second saddle point equation leads to the BCS equation fixing the $z_{j}(r)$ in each layer at $T=0$ and TF approximation as:

\begin{equation}
z_{j}=\frac{\rho_{j} U}{4}-\frac{\Delta^{2}}{\rho_{j} U}=\frac{\rho_{j} U}{4}-\frac{E_{b}^{2D}}{2},
\end{equation}

where $E^{2D}_{b}$ is the binding energy of the molecule in each layer and we have defined the gap $\Delta$ as:

\begin{equation}
\Delta=\sqrt{\frac{E_{b}^{2D} \rho_{j} U}{2}}.
\end{equation}

This definition of the gap is consistent with the definition given in Ref. \cite{Noz85} for small number of atoms per lattice site, if we identify the binding energy as $E_{b}^{2D}= U$.  We also now define,  $\mu_{eff}=z_{j}-\mu= \dfrac{\Delta^{2}}{2E_{b}^{2D}}-\dfrac{E_{b}^{2D}}{2}-\mu$ as the effective chemical potential in each layer. We now need to calculate the terms in the effective action that couple the different layers since we want to study the current perpendicular to the layers in which the atoms are confined. The tunneling contributions in the effective action $S_{eff,j\rightarrow j+1}^{tunnel}$ can be treated perturbatively. The lowest order perturbative expansion of the action with respect to the tunneling part yields:

\begin{equation}
S_{eff,j\rightarrow j+1}^{tunnel}= -\int_{0}^{\hbar \beta} d \tau \int d \mathbf{x} ~ J_{j\rightarrow j+1} \rho_{j} \cos\left( \theta_{j+1}-\theta_{j}\right),
\end{equation}
where
\begin{equation}
J_{j\rightarrow j+1}=\frac{J^{2}}{\frac{\Delta^{2}}{2E_{b}^{2D}}+E^{2D}_{b}}.
\end{equation}

The binding energy of the molecule in each layer is given by \cite{Petrov01}

\begin{equation}
E^{2D}_{b}=0.583~\sqrt{s}~E_{R}~exp\left(\frac{\lambda}{\sqrt{2\pi}a_{s}s^{-1/4}} \right). 
\end{equation}

$\lambda$ is the wavelength of the laser light and $a_{s}$ is the scattering length.  Having obtained the saddle point effective action $S_{eff}^{0}$ that depends on $\theta_{j}(r)$ and $\rho_{j}(r)$, we can now derive the equations of motion for the variables $\theta_{j}(r)$ and $\rho_{j}(r)$ from saddle point effective action through the extremum conditions $\partial S_{eff}^{0}/\partial \theta_{j}=0$ and the number equation. This leads to the following hydrodynamic equations for $\rho_{j}(r)$ and $\theta_{j}(r)$:

\begin{equation}
\frac{\hbar}{2}\frac{\partial \rho_{j}}{\partial t}=-\frac{\hbar^{2}}{4m}\vec{\nabla}.\left(\rho_{j}\vec{\nabla}\theta_{j} \right)+J_{j,j-1} \rho_{j} \sin(\theta_{j}-\theta_{j-1})-J_{j+1,j} \rho_{j} \sin(\theta_{j+1}-\theta_{j}), 
\end{equation}

\begin{equation}
-\frac{\hbar}{2}\frac{\partial \theta_{j}}{\partial t}=\frac{\hbar^{2}}{8m}\left(\nabla \theta_{j} \right)^{2}+V_{ext}+\mu_{eff}-J_{j+1,j}\cos(\theta_{j+1}-\theta_{j})- J_{j-1,j} \cos(\theta_{j}-\theta_{j-1}).
\end{equation}

Note that in equations (25) and (26), we have not explictly written the spatial dependence of $\rho_{j}$ and $\theta_{j}$ since it is understood that they depend on $r$. In the next section our starting point will be the above hydrodynamic equations and solve the corresponding equations of motion for the density and velocity fluctuations.

\section{Multibranch Bogoliubov Spectrum}

The equation of state enters through the density dependent chemical potential. We assume the power-law form of the equation of state as $\mu_{eff}(\rho)=C\rho^{\gamma}$ \cite{Ghosh06}. $\gamma$ is an effective polytropic index. The polytropic approximation has the advantage of allowing one to get analytical expressions for the eigenfunctions and eigenfrequencies of collective modes for various superfluid regimes in a unified way. At equilibrium, the two dimensional density profile takes the form at each layer $\rho_{0}(r)=\left(\frac{\mu_{eff}}{C} \right)^{1/\gamma}\left(1-\tilde{r}^{2} \right)^{1/\gamma} $, where $\tilde{r}=\frac{r}{R}$ and $R=\sqrt{\dfrac{2\mu_{eff}}{m\omega_{r}^{2}}}$. Linearizing around equilibrium, $\rho_{j}=\rho_{0}+\delta \rho_{j}$, $\theta_{j}=\delta \theta_{j}$ and $\mu_{eff}(\rho_{j})=\mu_{eff}(\rho_{0})+\dfrac{\partial \mu_{eff}}{\partial \rho_{j}}|_{\rho=\rho_{0}}\delta \rho_{j}$. The equations of motion for the density and phase fluctuations are

\begin{equation}
\frac{\hbar}{2}\frac{\partial \delta \rho_{j}}{\partial t}=-\frac{\hbar^{2}}{4m}\vec{\nabla}.\left(\rho_{0}\vec{\nabla}\delta \theta_{j} \right)+\frac{J^{2}\rho_{0}}{\dfrac{\Delta^{2}}{E_{b}^{2D}}+E^{2D}_{b}}\left(2\delta \theta_{j}-\delta \theta_{j+1}-\delta \theta_{j-1} \right),  
\end{equation}

\begin{equation}
\frac{\hbar}{2}\frac{\partial \delta \theta_{j}}{\partial t}=-\frac{\partial \mu_{eff}}{\delta \rho_{j}}|_{\rho_{j}=\rho_{0}}\delta \rho_{j}.
\end{equation}

In deriving the above equations,we have assumed that the optical lattice is deep so that $J_{j\pm 1,j}<<\mu_{eff}$. The second order equation of motion for the density fluctuation is given by

\begin{equation}
\frac{\hbar}{2}\frac{\partial^{2} \delta \rho_{j}}{\partial t^{2}}=\frac{\hbar}{8m}\vec{\nabla}.\left(\rho_{0}\vec{\nabla}\frac{\partial \mu_{eff}}{\partial \rho_{j}}|_{\rho_{j}=\rho_{0}}\delta \rho_{j} \right)+\frac{2C\gamma \rho_{0}^{\gamma}F}{\hbar}\left(-2\delta \rho_{j}+\delta \rho_{j+1}+\delta \rho_{j-1} \right),  
\end{equation}

where 

\begin{equation}
F=\frac{J^{2}}{\left(\frac{\Delta^{2}}{E_{b}^{2D}}+E_{b}^{2D} \right) }
\end{equation}

A further simplification yields

\begin{equation}
\frac{\partial^{2} \delta \rho_{j}(\tilde {r})}{\partial t^{2}}=\frac{\mu_{eff} \gamma}{mR^{2}}\vec{\nabla}_{\tilde{r}}.\left((1-\tilde{r}^{2})^{1/\gamma}\vec{\nabla}_{\tilde{r}}(1-\tilde{r}^{2})^{1-1/\gamma}\delta \rho_{j} (\tilde{r}) \right)+\frac{4C\gamma \rho_{0}^{\gamma}(\tilde{r})F}{\hbar^{2}}\left(-2\delta \rho_{j}(\tilde{r})+\delta \rho_{j+1}(\tilde{r})+\delta \rho_{j-1}(\tilde{r}) \right).  
\end{equation}

We assume a normal mode solution of the form

\begin{equation}
\delta \rho_{j}(\tilde{r},z,t)=\delta \rho_{0}(\tilde{r})~exp~i\left(\omega_{\alpha}(k)t-jkd \right). 
\end{equation}

This yields
\begin{equation}
-\tilde{\omega}^{2}_{\alpha}\delta \rho_{0}(\tilde{r})=\frac{\gamma}{2}\vec{\nabla}_{\tilde{r}}.\left((1-\tilde{r}^{2})^{1/\gamma}\vec{\nabla}_{\tilde{r}}(1-\tilde{r}^{2})^{1-1/\gamma}\delta \rho_{j} (\tilde{r})\right) -\frac{16\gamma C \rho_{0}^{\gamma}(\tilde{r})F}{E_{R}^{2}}\sin^{2}\left(\frac{kd}{2} \right)\delta \rho_{0}(\tilde{r}), 
\end{equation}

where, $\tilde \omega=\dfrac{\omega}{\omega_{R}}$. Here $\alpha$ is a set of two quantum numbers: radial quantum number $n_{r}$ and the angular quantum number $m$.

For $k=0$, it reduces to a two-dimensional eigenvalue problem and the solutions of it can be obtained analytically \cite{Ghosh06} by generalizing the method of \cite{Pethick02} to any polytropic equation of state. The energy spectrum is given by

\begin{equation}
\tilde{\omega}^{2}_{\alpha}=|m|+2n_{r}\left(\gamma (n_{r}+|m|)+1 \right) 
\end{equation}

The corresponding normalized eigenfunction is given by
\begin{equation}
\delta \rho_{\alpha}=A\left(1-\tilde{r}^{2} \right)^{(1/\gamma-1)}\tilde{r}^{|m|}P_{n_{r}}^{(1/\gamma-1,|m|)}(2\tilde{r}^{2}-1)~exp\left(im\phi \right)  
\end{equation}

where $P_{n}^{a,b}(x)$ is a Jacobi polynomial of order $n$ and $\phi$ is the polar angle. Also, the normalization constant $A$ is given by

\begin{equation}
A^2 = \frac{2^{2-2/\gamma}}{\sqrt{\pi} R^2}
\frac{[\Gamma(n_r+1)]^2 \Gamma(1/\gamma)\Gamma(2/\gamma + 2 n_r + |m|)}
{\Gamma(1/\gamma-1/2)[\Gamma(1/\gamma + n_r)]^2 \Gamma(2n_r + |m| +1)}.
\end{equation}

For $k\not= 0$, we expand the density fluctuation as

\begin{equation}
\delta \rho_{0} = \sum_{\alpha} b_{\alpha} \delta \rho_{\alpha} (\tilde{r},\phi).
\end{equation}

Substituting the above expansion into equation (33), we obtain

\begin{eqnarray}
0 & = & [\tilde \omega_{\alpha}^2 - [|m| + 2 n_r ( \gamma (n_r +|m|) +1)]
\nonumber \\ & - &
B_{0}\sin^{2}\left(\frac{kd}{2} \right) ]  b_{\alpha} +
 B_{0}\sin^{2}\left(\frac{kd}{2} \right) \sum_{\alpha^{\prime}} M_{\alpha \alpha^{\prime}} 
b_{\alpha^{\prime}}.
\end{eqnarray}

where 

\begin{equation}
B_{0}=\frac{16\gamma \mu_{eff} F}{E_{R}^{2}}
\end{equation}

The matrix element $M_{\alpha,\alpha ^{\prime}}$

\begin{eqnarray}
M_{\alpha \alpha^{\prime}} & = & \tilde{A}^2 \int d^2 \tilde r
(1-\tilde r^2)^{2\gamma_0} \tilde r^{2+|m| + |m^{\prime}|} e^{i(m - m^{\prime}) \phi}
\nonumber \\ & \times & P_{n_r^{\prime}}^{(\gamma_0,|m^{\prime}|)}(2 \tilde r^2-1)
P_{n_r}^{(\gamma_0,|m|)}(2 \tilde r^2-1),
\end{eqnarray}

where $\gamma_{0}=1/\gamma -1$, $\tilde{A}^{2}=\pi R^{2}A^{2}$. Equation (40) is the central result of this work, which is derived for the first time.

\begin{figure}[t]
\hspace{-1.5cm}
\includegraphics{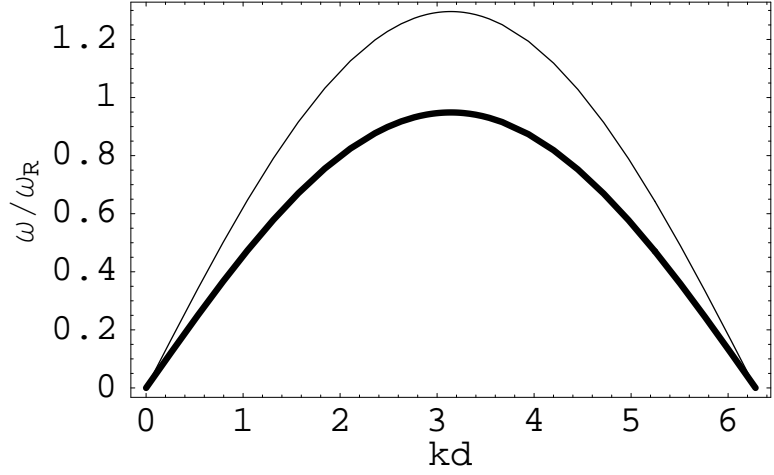} 
\caption{Plots of the phonon modes ($m=0, n_{r}=0$) in the BEC regime (thick line,$\dfrac{J}{E_{R}}=1, \gamma=0.8, \Delta=2E_{F},E_{bec}^{2D}=8E_{F}$ ) and BCS regime (thin line, $E_{F}=25E_{bcs}^{2D}, \dfrac{J}{E_{R}}=1, \gamma=0.6, \Delta=0.4E_{F}$).}
\label{fig:figure_1}
\end{figure}

 Before proceeding ahead, let us have a look at typical experimental values. Let us now consider the experiment of Greiner et al. \cite{Greiner03}. They had prepared ultracold gas of fermionic $^{40}K$ atoms in the lowest energy spin states $|f=9/2, m_{f}=-7/2 \rangle $ and $|f=9/2, m_{f}=-9/2 \rangle $, where $f$ is the total atomic angular momentum and $m_{f}$ the magnetic quantum number. The Feshbach resonance occurs at a magnetic field $B_{0}=202.1\pm 0.1~G$ and has a width of $w=7.8\pm 0.6~G$. When $B<B_{0}$, the $s$-wave scattering length $a_{s}$ is positive and a weakly bound molecular BEC is formed. At $B=B_{0}$ (unitary limit), $a_{s}\rightarrow \pm \infty$, which corresponds to a very small binding energy ( see equation 24). Beyond the resonance ($B>B_{0}$), $a_{s}$ is negative and the binding energies in the BCS($E_{bcs}^{2D}$),unitary($E_{uni}^{2D}$) and the BEC($E_{bec}^{2D}$) region are related as  $E_{bcs}^{2D}<E_{uni}^{2D}<E_{bec}^{2D}$. The scattering length $a_{s}$ is determined as $a_{s}=174 a_{0}\left(1+\dfrac{w}{(B_{0}-B)} \right) $ \cite{Greiner03}, where $a_{0}$ is the Bohr radius. 

Since we are interested in the crossover, we will focus our attention near the unitary limit. On the BEC side of the unitary limit, if we take $B=201.2~G<B_{0}$, we find that the scattering length on the BEC side is $a_{s}^{bec}\approx 3.78\times 10^{-7}m$ and $E_{bec}^{2D}\approx 4.26 E_{R}$. In the BEC region, the chemical potential is $\mu=-E_{bec}^{2D}/2$, $\Delta >> E_{F}$ and $E_{F}<E_{bec}^{2D}$. Using these values, we find that $E_{bec}^{2D}>>\dfrac{\Delta^{2}}{E_{bec}^{2D}}$. On the BCS side near the unitary limit, $B=202.2~G$, $a_{s}^{bcs}\approx -3.60\times 10^{-7}m$ and $E_{bcs}^{2D}\approx 0.22 E_{R}$, $\mu=E_{F}$(Fermi energy), $\Delta \rightarrow 0$, $E_{F}>>E_{bcs}^{2D}$.

To calculate the multibranch Bogoliubov spectrum, we need to know the adiabatic index $\gamma$. $\gamma=2/3$ denotes the unitary as well as the BCS limit\cite{Parish05}. For the BEC side of the unitary limit $\gamma>2/3$ and for the BCS side of the unitary limit we take $0.6<\gamma<2/3$. 

In \figurename~\ref{fig:figure_1}, we plot the lowest ($n_{r}=0,m=0$) mode for the BEC side of unitary limit (thick line, $\dfrac{J}{E_{R}}=1,\gamma=0.8, \Delta=2E_{F}, E_{bec}^{2D}=8E_{F}$) and BCS side of unitary limit (thin line,$E_{F}=25E_{bcs}^{2D},\dfrac{J}{E_{R}}=1,\gamma=0.6,\Delta=0.4E_{F}$).Clearly, we find that in the entire Brillioun zone, the frequency of the phonon mode in the BCS side of unitary limit is greater than that in the BEC side of unitary limit. In the limit of long wavelength, the $n_{r}=0$ mode is phonon like. On the BEC side $\omega_{bec} \approx k \sqrt{\dfrac{\gamma (2-\gamma) \mu_{eff}}{m_{bec}^{*}}}$, where $ m_{bec}^{*}=\dfrac{E_{bec}^{2D} \hbar^{2}}{2J^{2}d^{2}}$ is the effective mass in the BEC side of unitary limit. On the BCS side the low wavelength behaviour is $\omega_{bcs} \approx k \sqrt{\dfrac{\gamma (2-\gamma) \mu_{eff}}{m_{bcs}^{*}}}$, where $ m_{bcs}^{*}=\dfrac{\hbar^{2}\Delta^{2}}{2J^{2} d^{2}E_{bcs}^{2D}}$. In Ref. \cite{Pitae05}, it was found that the effective mass increases as the density of the particle increases. This observation is in accordance with our result, if we substitute $\Delta^{2}/E_{bcs}^{2D}$ from equation (21) in the expression for $m_{bcs}^{*}$. Note that the expression for $m_{bec}^{*}$ coincides with that found in \cite {Noz85} and $m_{bec}^{*}>m_{bcs}^{*}$. By gradually changing the magnetic field when we go from the BEC regime to the BCS regime, we observe an increase in the phonon mode which is expected for a collisionless Fermi gas, where the elastic collision rate is strongly reduced by Pauli blocking. In the absence of coupling between the axial modes with the density along the radial direction (for the $n_{r}=0$ mode),the effect of Pauli blocking is strong and $\omega_{bec}<\omega_{bcs}$ in the entire Brillioun zone. the coupling between the axial modes and the radial density enhances the elastic collision rate.

\begin{figure}[t]
\hspace{-1.5cm}
\includegraphics{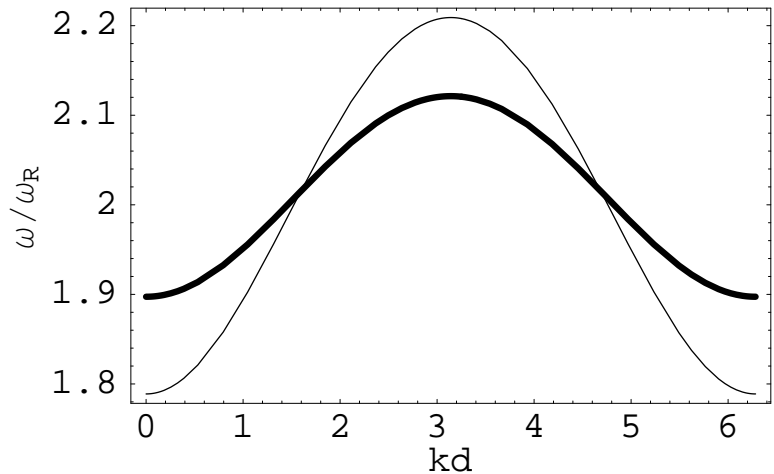} 
\caption{Plots of the monopole modes ($m=0, n_{r}=1$) in the BEC regime (thick line,$\dfrac{J}{E_{R}}=1, \gamma=0.8, \Delta=2E_{F}, E_{bec}^{2D}=4E_{F}$ ) and BCS regime (thin line, $E_{F}=25E_{bcs}^{2D}, \dfrac{J}{E_{R}}=1, \gamma=0.6, \Delta=0.4E_{F}$). One can clearly see the difference in the long wavelength and the short wavelength behaviour.}
\label{fig:figure_2}
\end{figure}

In \figurename~\ref{fig:figure_2}, we show the monopole modes ($m=0,n_{r}=1$)in the BEC side of unitary limit (thick line) and the BCS side of unitary limit (thin line).The monopole modes show a very peculiar behaviour. In the long wavelength region, $\omega_{bec}>\omega_{bcs}$. As we move away from the center of the Brillioun zone, there is a cross over and $\omega_{bec}<\omega_{bcs}$. If we look at equation (21), we find that in the long wavelength region, the term $2n_{r}[\gamma (n_{r}+1)]$ (this part indicates the coupling between the radial density and the axial frequencies and is finite for $n_{r}\not=0$)dominates over the part determined by the optical lattice $ \left(1-\sum_{\alpha} M_{\alpha,\alpha} \right) B_{0}\dfrac{k^{2}d^{2}}{4} $. Consequently, the elastic damping rate is high and $\omega_{bec}>\omega_{bcs}$ as expected from experiments on elongated Fermi gases \cite{Bartenstein04}, because the term $2n_{r}[\gamma (n_{r}+1)]$ is greater in the BEC region as compared to that in the BCS region (since $\gamma_{bec}>\gamma_{bcs}$). On moving away from the center of the Brillioun zone, the term proportional to $\sin^{2}\left( \dfrac{kd}{2}\right) $ starts dominating and as we go from the BEC to BCS regime, we probably enter the collisionless phase and the elastic collision rate decreases and as a result $\omega_{bec}<\omega_{bcs}$. Note that Pauli blocking reduces the binding energy because of which $E_{bcs}^{2D}<E_{bec}^{2D}$.  For both the phonon modes and the monopole modes, the frequency ($\omega_{uni}$) in the unitary limit($\gamma=2/3$) will lie between $\omega_{bec}$ and $\omega_{bcs}$. Eventhough $\gamma=2/3$ in both the unitary and BCS limit, $\omega_{uni}<\omega_{bcs}$ because from equation (21) we notice that $E_{uni}^{2D}>E_{bcs}^{2D}$. Our analysis of the low energy Bogoliubov modes in the BCS-BEC crossover was obtained in the Thomas-Fermi approximation, which is valid only in the large density (weak coupling) regime. The important excitations in the BCS limit involve broken pairs. With increaing attraction these are pushed to very high energies, and in the Bose limit it is the collective modes which are the dominant low-energy excitations. Thus a proper description of the intermediate coupling regime must include both broken pairs and the collective modes. It is clear that present approach is not suitable to describe the intermediate as well as well as the strong coupling regime. A better method is the LDA (local density approximation) which is known to give satisfactory results in the intermediate as well as the strong coupling regime \cite{Randeria95}. The spectra of the phonon and the monopole modes in the different regimes can be observed in the Bragg scattering experiments as these spectra have been observed in Ref. \cite{Stein03} for weakly interacting BEC. By measuring the sound velocity in pulse propagation experiments and by observing the low-energy Bogoliubov spectrum in the Bragg spectroscopy, one can make a clear identification of various superfluid regimes along the BCS-BEC crossover. The results presented in this work may be useful for guiding experiments, which look for signatures of BEC-BCS crossover in optical lattices.

\section{Conclusions}
We have studied the Bogoliubov spectrum of an elongated Fermi superfluid confined in an one-dimensional superfluid along the Bose-Einstein-condensate (BEC)-Bardeen-Cooper-Schrieffer (BCS) crossover. Using the hydrodynamic approach, we have analytically calculated the effective mass and the multibranch Bogoliubov spectrum in the BEC and BCS side of the unitary limit. We have shown that the effective mass increases as the system crosses from the BCS side to the BEC side. The Bogoliubov axial excitation frequencies on either side of the unitary limit show a strong dependence on the coupling with the radial density and the binding energy and thus provide valuable information on the physical behavior of the system. The frequency of the phonon mode in the BCS side is greater than that in the BEC side. On the other hand near the center of the Brillioun zone, we show that the monopole frequency on the BEC side is greater than that on the BCS side but as we go towards the edge of the Brillioun zone, the monopole frequency on the BCS side becomes greater than that on the BEC side. The various Bogoliubov frequencies calculated here can be measured by Bragg scattering experiments. 

\begin{acknowledgments}
The author is grateful to the Max Planck Institute for Physics of Complex Systems, Dresden, Germany for the hospitality and for providing the facilities for carrying out the present work. I am grateful to Bijaya Sahoo for some useful discussions.
\end{acknowledgments}


\begin{thebibliography}{99}

\bibitem{Modugno03}
G. Madugno et al., 
Phys. Rev. A {\textbf{68}}, 011601(R) (2003).  
%
\bibitem{Jochim03}
S. Jochim et al., Phys. Rev. Letts. {\textbf{91}}, 240402 (2003).
%
\bibitem{Hofstter02}
W. Hofstetter et al., Phys. Rev. Letts. {\textbf{89}}, 220407 (2002).
%
\bibitem{Orso04}
G. Orso, L .P. Pitaevskii and S. Stringari, Phys. Rev. Letts. {\textbf{93}},020404 (2004).
%
\bibitem{Pitae05}
L.P. Pitaevskii, S. Stringari and G. Orso, Phys. Rev. A, {\textbf{71}},053602 (2005).
%
\bibitem{Parish05}
M. M. Parish, B. Mihaila, E. M. Timmermans, K. B. Blagoev and P. B. Littlewood, 
Phys. Rev. B {\textbf{71}}, 064513 (2005), H. Heiselberg, New J. Phys. {\textbf{6}}, 137 (2004),H. Heiselberg, Phys. Rev. A {\textbf{73}}, 013607 (2006), C. Chin, Phys. Rev. A {\textbf{72}}, 041601(R) (2005), D. E. Sheehy and L. Radzihovsky, Phys. Rev. Letts. {\textbf{96}}, 060401 (2006).
%
\bibitem{Ghosh06}
T. Ghosh and K. Machida, Phys. Rev. A {\textbf{73}}, 013613 (2006), H. Heiselberg, Phys. Rev. Letts. {\textbf{93}}, 040402 (2004), R. Comescot, M. Y. Kagan and S. Stringari, Phys. Rev. A {\textbf{74}}, 042717 (2006), J. Yin, Y-Li. Ma and G. Huang, Phys. Rev. A {\textbf{74}}, 013609 (2006), Y. Zhou and G. Huang, Phys. Rev. A {\textbf{75}}, 023611 (2007), S. Stringari, Europhysics Letts. {\textbf{65}}, 749 (2004), H. Hu, A. Minguzzi, X. J. Liu and M. P. Tosi, Phys. Rev. Letts. {\textbf{93}}, 190403 (2004), Y. E. Kim and A. L. Zubarev, Phys. Rev. A {\textbf{72}}, 011603(R) (2005),N. Manini and L. Salasnich, Phys. Rev. A {\textbf{71}}, 033625 (2005), M. Holland, J. Park and R. Walser, Phys. Rev. Letts. {\textbf{86}}, 1915 (2000), T. N. Silva and E. J. Mueller, Phys. Rev. A {\textbf{72}}, 063614 (2004).
%
\bibitem{Bartenstein04}
M. Bartenstein et al., Phys. Rev. Letts. {\textbf{92}}, 203201 (2004), J. Kinast et al., Phys. Rev. Letts. {\textbf{92}}, 150402 (2004).
%
\bibitem{Houbiers98}
M. Houbiers et al., Phys. Rev. A, {\textbf{57}}, R1497 (1998), W. C. Stwalley, Phys. Rev. Letts. {\textbf{37}}, 1628, (1976), E. Tiesinga, B. J. Verhaar and H. T. C. Stoof, Phys. Rev. A, {\textbf{47}}, 4114 (1993).
%
\bibitem{Hara02}
K. M. O'Hara et al., Science {\textbf{298}}, 217 (2002), C. A. Regal et al., Phys. Rev. Letts. {\textbf{92}}, 040403, (2004), M. Bartenstein et al., Phys. Rev. Letts. {\textbf{92}}, 120401 (2004), T. Bourdel et al., Phys. Rev. Letts. {\textbf{93}}, 050401 (2004),C. Chin et al., Science {\textbf{305}}, 1128 (2004).
%
\bibitem{Wouters04}
M. Wouters, J. Tempere and J. T. Devreese, Phys. Rev. A, {\textbf{70}}, 013616 (2004).
%
\bibitem{Martikainen03}
J.-P. Martikainen, H. T. C. Stoof, Phys. Rev. A {\textbf{68}}, 013610 (2003).
%
\bibitem{Noz85}
P. Nozieres and S. Schmitt-Rink, J. low Temp. Phys. {\textbf{59}}, 195 (1985).
\bibitem{Petrov01}
%
D. S. Petrov and G. V. Shlyapnikov, Phys. Rev. A {\textbf{64}}, 012706 (2001). 
%
\bibitem{Palo98}
S. De Palo, C. Castellani and C. Di Castro, Phys. Rev. B {\textbf{60}}, 564 (1998).
%
\bibitem{Pethick02}
C.J. Pethick and H. Smith, Bose Einstein Condensation in Dilue Gases (Cambridge University Press, Cambridge, New York, 2002), p.181.
%
\bibitem{Greiner03}
M. Greiner, C. A. Regal and D. S. Jin, Nature {\textbf{426}}, 537 (2003).
%
\bibitem{Randeria95}
M. Randeria in Bose Einstein Condensation (Cambridge University Press, Cambridge, U.K., 1995).
%
\bibitem{Stein03}
J. Steinhauer et. al., Phys. Rev. Letts. {\textbf{90}}, 060404, (2003).


\end{thebibliography}
\end{document}